\documentclass[12pt]{article}
\usepackage{graphicx}
\usepackage{amsmath}
\usepackage{amsfonts}
\usepackage{amssymb}
\usepackage{bbm}
\usepackage{hyperref}

\begin{document}

\newcommand{\ba}[1]{\begin{array}{#1}} \newcommand{\ea}{\end{array}}
\newcommand{\cleqn}{\setcounter{equation}{0}}

\numberwithin{equation}{section}

%%%%%%%%%%%% BORUT MACROS %%%%%%%%%%%%%%%%%%%

% A useful Journal macro
\def\Journal#1#2#3#4{{#1} {\bf #2}, #3 (#4)}

% Some useful journal names
\def\NCA{\em Nuovo Cimento}
\def\NIM{\em Nucl. Instrum. Methods}
\def\NIMA{{\em Nucl. Instrum. Methods} A}
\def\NPB{{\em Nucl. Phys.} B}
\def\PLB{{\em Phys. Lett.}  B}
\def\PRL{\em Phys. Rev. Lett.}
\def\PRD{{\em Phys. Rev.} D}
\def\ZPC{{\em Z. Phys.} C}

% Some other macros used in the sample text
\def\st{\scriptstyle}
\def\sst{\scriptscriptstyle}
\def\mco{\multicolumn}
\def\epp{\epsilon^{\prime}}
\def\vep{\varepsilon}
\def\ra{\rightarrow}
\def\ppg{\pi^+\pi^-\gamma}
\def\vp{{\bf p}}
\def\ko{K^0}
\def\kb{\bar{K^0}}
\def\al{\alpha}
\def\ab{\bar{\alpha}}
%\def\CPbar{\hbox{{\rm CP}\hskip-1.80em{/}}}%temp replacemt due to no font

%%%%%%%%%%%%%%%%%%%%%%%%%%%%%%%%%%%%%%%%%%%%%%%%%%%%%%%%%%
\def\np{Nucl. Phys. {\bf B}}\def\pl{Phys. Lett. {\bf B}}
\def\mpl{Mod. Phys. {\bf A}}\def\ijmp{Int. J. Mod. Phys. {\bf A}}
\def\cmp{Comm. Math. Phys.}\def\prd{Phys. Rev. {\bf D}}

\def\oa{\bigcirc\!\!\!\! a}
\def\ob{\bigcirc\!\!\!\! b}
\def\oc{\bigcirc\!\!\!\! c}
\def\oi{\bigcirc\!\!\!\! i}
\def\oj{\bigcirc\!\!\!\! j}
\def\ok{\bigcirc\!\!\!\! k}
\def\ve{\vec e}\def\vk{\vec k}\def\vn{\vec n}\def\vp{\vec p}
\def\vr{\vec r}\def\vs{\vec s}\def\vt{\vec t}\def\vu{\vec u}
\def\vv{\vec v}\def\vx{\vec x}\def\vy{\vec y}\def\vz{\vec z}

\def\ve{\vec e}\def\vk{\vec k}\def\vn{\vec n}\def\vp{\vec p}
\def\vr{\vec r}\def\vs{\vec s}\def\vt{\vec t}\def\vu{\vec u}
\def\vv{\vec v}\def\vx{\vec x}\def\vy{\vec y}\def\vz{\vec z}

\newcommand{\AdS}{\mathrm{AdS}}
\newcommand{\dd}{\mathrm{d}}
\newcommand{\eee}{\mathrm{e}}
\newcommand{\sgn}{\mathop{\mathrm{sgn}}}

\def\a{\alpha}
\def\b{\beta}
\def\g{\gamma}

%%%%%%%%%%%%%%%%%%%  BORUT MACROS %%%%%%%%%%%%%%%%%%%%%%%%%%%%%
\newcommand\lsim{\mathrel{\rlap{\lower4pt\hbox{\hskip1pt$\sim$}}
    \raise1pt\hbox{$<$}}}
\newcommand\gsim{\mathrel{\rlap{\lower4pt\hbox{\hskip1pt$\sim$}}
    \raise1pt\hbox{$>$}}}

\newcommand{\beq}{\begin{equation}}
\newcommand{\eeq}{\end{equation}}
\newcommand{\bea}{\begin{eqnarray}}
\newcommand{\eea}{\end{eqnarray}}
\newcommand{\bem}{\begin{pmatrix}}
\newcommand{\eem}{\end{pmatrix}}
\newcommand{\noi}{\noindent}

%%%%%%%%%%%%%%%%%%%%%%%%%%%%%%%%%%%%%%%%%%%%%%%%%%%%%%%%%%%%%%%%
%%%%%%%%%%%%%%%%%%%%%%%%%%%%%%%%%%%%%%%%%%%%%%%%%%%%%%%%%%%%%%%%%%%

\begin{flushright}
April, 2013
\end{flushright}

\bigskip

\begin{center}

{\Large\bf On the matching method and the Goldstone theorem in holography}
\vspace{1cm}

\centerline{Borut Bajc$^{a,b,}$\footnote{borut.bajc@ijs.si}, and Adri\'{a}n R. Lugo$^{a,c,}
$\footnote{lugo@fisica.unlp.edu.ar}}

\vspace{0.5cm}
\centerline{$^{a}$ {\it\small J.\ Stefan Institute, 1000 Ljubljana, Slovenia}}
\centerline{$^{b}$ {\it\small Department of Physics, University of Ljubljana, 1000 Ljubljana, Slovenia}}
\centerline{$^{c}$ {\it\small Departamento de F\'\i sica and IFLP-CONICET, }}
\centerline{ {\it\small Facultad de Ciencias Exactas, Universidad Nacional de La Plata,}}
\centerline{ {\it\small  C.C. 67, 1900 La Plata, Argentina}}

\end{center}

\bigskip

\begin{abstract}
We study the transition of a scalar field in a fixed $AdS_{d+1}$ background between an 
extremum and a minimum of a potential. We compute analytically the solution to the 
perturbation equation for the vev deformation case by generalizing the usual matching method to higher orders 
and find the propagator of the boundary theory operator defined through the AdS-CFT correspondence. 
We show that, contrary to what happens at the leading order of the matching method, the next-to-leading 
order presents a simple pole at $q^2=0$ in accordance with the 
Goldstone theorem applied to a spontaneously broken dilatation invariance.
 
\end{abstract}

\clearpage

\tableofcontents

%%%%%%%%%%%%%%%%%%%%%%%%%%%%\section{Introduction}%%%%%%%%%%%%%%%%%%%%%%%%%%%%

\section{Introduction and results}

The gravity plus real scalar field system in asymptotic anti de Sitter (AdS) spaces is the simplest playground for
the AdS-CFT correspondence \cite{Maldacena:1997re,Gubser:1998bc,Witten:1998qj}. It is relatively easy to solve
and it gives some insight into what the correspondence is and what it means (for a partial list see \cite{Girardello:1998pd,Freedman:1999gk,Arutyunov:2000rq,Muck:2001cy,Martelli:2001tu,Berg:2002hy,Freedman:2003ax}).
Unfortunately several simple statements get a bit obscured by technical details and the complexity of the gravity
system. Since one would expect that weakly coupled gravity has a smooth non-interacting limit\footnote{In particular
cases there could be subtle exceptions though, see for example \cite{Cvetic:1992bf}. }, it may be useful to study the
no back-reaction limit of this system: a real scalar field in a fixed AdS background, and no gravity at all (for some
reviews which treat the subject see for example \cite{Aharony:1999ti,DeWolfe:2000xi,D'Hoker:2002aw,Skenderis:2002wp}).
It has been shown long ago \cite{Bianchi:2001de} that a potential of the real scalar field presenting a UV maximum and a
IR minimum with a nontrivial solution (vev deformation) for the scalar field leads to a 2-point scalar correlator on the
boundary with a simple $1/q^2$ pole for small $q$. This is the consequence of the spontaneously broken dilatation
invariance and the massless state is the Nambu-Goldstone boson. Since then such a behavior has been re-derived on
several occasions, for example in \cite{Bajc:2012vk}, where we considered a piece-wise quadratic potential in order to 
allow analytic treatment. In spite of the generality of the argument, the conclusion has been put into question recently in
\cite{Hoyos:2012xc}, where a $1/q^{2\nu_{IR}}$ propagator with $\nu_{IR}>d/2$ has been explicitly found in the
$q\to 0$ limit. The presence of this term is expected: it is an unavoidable part of the continuous spectrum. 
What is surprising is that this was found to be the leading order in the limit $q\to 0$. 

The purpose of this paper is to generalize the piece-wise quadratic potential of \cite{Bajc:2012vk} to more
general ones with the main goal to clarify which is the leading low energy behavior of the propagator, the (expected on
general grounds) $1/q^2$ of \cite{Bianchi:2001de}, \cite{Bajc:2012vk} or the (surprising) $1/q^{2\nu_{IR}}$ of
\cite{Hoyos:2012xc}. Although from \cite{Bianchi:2001de} we expect it to be the $1/q^2$ (as we will confirm), it will
be interesting to see what went wrong with the method or the approximation used in \cite{Hoyos:2012xc}. 
Again, we are here talking only about the leading behavior for small $q$: in the complete result both 
$1/q^2$ and $1/q^{2\nu_{IR}}$ are present. In this respect the result of \cite{Hoyos:2012xc} reproduces 
part of the unavoidable continuous spectrum of the 2-point correlator. What we will show here is 
that on top of that there is also a discrete state at $q^2=0$. The bulk state dual to this pole is the zero mode 
described in ref. \cite{Berg:2006xy}, which is normalizable in vev flows. 

In particular, we show that a consistent treatment of the matching method requires a next-to-leading order calculation.
The result confirms what we found in \cite{Bajc:2012vk}: the propagator of the boundary theory defined through the 
AdS-CFT correspondence for the vev deformation case (i.e. for the spontaneous symmetry breaking case or, equivalently, 
for the case of a non-zero vev of the boundary operator for vanishing source \cite{Skenderis:2002wp}) 
in the gravity no-back-reaction limit
has a simple pole at $q^2=0$, signaling the presence of a Goldstone mode.
The argument can be summarized as follows.
The leading order solution for the perturbation in a small $q$ expansion is
given by
\beq
\label{xileading}
\xi(z;q)\approx C_+(q)\;\xi_+(z)+C_-(q)\;\xi_-(z)
\eeq
with the limits,
\bea
\xi_+(z)\xrightarrow{z\to 0} a_+^{UV}\;z^{\Delta^{UV}}\qquad&;&\qquad
\xi_+(z)\xrightarrow{z\to\infty} a_+^{IR}\;z^{d-\Delta^{IR}}\\
\xi_-(z)\xrightarrow{z\to 0} a_-^{UV}\;z^{d-\Delta^{UV}} \qquad&;& \qquad
\xi_-(z)\xrightarrow{z\to\infty} a_-^{IR}\;z^{\Delta^{IR}}
\eea
where $\Delta^{UV}>d/2$ and $\Delta^{IR}>d$ are defined in (\ref{Delta}), while $d$ is the boundary space-time dimension. 
There is only one relevant correction to (\ref{xileading}) for $q\to0$, leading to
\beq
\xi(z;q)\approx C_+(q)\;\left(\xi_+(z)+q^2\;\delta\xi_+(z)\right)+C_-(q)\;\xi_-(z)
\eeq
This next-to-leading order correction $\delta\xi_+(z)$ behaves in the limiting cases as
%\footnote{
%We keep here the same apparently obscure notation as in the body of the paper: it will become clear later on.
%}
\beq
\delta\xi_+(z)\xrightarrow{z\to0/\infty}\epsilon_{++}^{UV/IR}\;\xi_-(z)
\eeq
The matching with the (conveniently normalized) Bessel K solution for $z\to\infty$
is done at fixed, small $q\,z\,$:
\beq
C_+(q)\;a_+^{IR}\;z^{d-\Delta^{IR}}+\left(C_-(q)+q^2\;C_+(q)\;\epsilon_{++}^{IR}\right)\;
a_-^{IR}\;z^{\Delta^{IR}}\approx z^{d-\Delta^{IR}} +
\gamma\;q^{2\nu_{IR}}\;z^{\Delta^{IR}}
\eeq
($\gamma$ is given in (\ref{gamma})) with the result for $\nu_{IR}>1$ (always true for $d>2$)
\beq
C_+(q)=1/a_+^{IR}\qquad,\qquad C_-(q)=-C_+(q)\;\epsilon_{++}^{IR}\;q^2
\eeq
On the other side the boundary propagator is found according to the AdS/CFT prescription
at the opposite limit $z\to0$:
\beq
C_+(q)\;a_+^{UV}\;z^{\Delta^{UV}}+\left(C_-(q)+q^2\;C_+(q)\;\epsilon_{++}^{UV}\right)\;
a_-^{UV}\;z^{d-\Delta^{UV}}
%\equiv C_+(q)\;a_+^{UV}
\equiv\frac{a_+^{UV}}{a_+^{IR}}\;\left(z^{\Delta^{UV}}+\frac{z^{d-\Delta^{UV}}}{G_2(q)}\right)
\eeq
from where for $q\to0$
\beq
G_2(q)=\frac{a_+^{UV}/a_-^{UV}}{\epsilon_{++}^{UV}-\epsilon_{++}^{IR}}\times\frac{1}{q^2}
\eeq
follows.
Had we stopped at the formally leading order (i.e. without $\delta\xi_+(z)$ or $\epsilon^{UV}_{++}$), we
would have got a completely different (and wrong) propagator in the $q\to0$ limit:
\beq
G_2(q)\to\left(\frac{a_+^{UV}}{a_-^{UV}}\right)\left(\frac{a_-^{IR}}{a_+^{IR}}\right)\frac{1}{\gamma}\times q^{-2\nu_{IR}}
\eeq
This is the essence of the argument, all details will be given in the paper.

Our philosophy in obtaining these results will be similar to that in \cite{Bajc:2012vk}, which has become usual in
applications of the AdS/CFT duality to condensed matter systems (see for example \cite{Hartnoll:2009sz}): we
do not know which the dual Lagrangian on the boundary is, but we simply define the boundary theory from the
bulk theory through the holographic dictionary, hoping that it gives sensible answers and is consistent with the
usual quantum field theory rules.

The plan of the paper is the following.
After setting the notation and main formulae in Section \ref{system},
we first show in Section \ref{BPS} that even in this no-backreaction limit one can get a 
BPS-type solution to the first order equation of motion provided that the potential is written in 
terms of a properly defined superpotential. 
The most important part of the paper is Section \ref{methods}, where we describe 
in details the matching method, showing that the method is correct, but the leading order is
not enough for this issue.

%%%%%%%%%%%%%%%%%%%%%%%%%%%%
\section{Preliminaries}
%%%%%%%%%%%%%%%%%%%%%%%%%%%%

%%%%%%%%%%%%%%%%%%%%%%%%%%%%
\subsection{\label{system}The system}
%%%%%%%%%%%%%%%%%%%%%%%%%%%%

We consider a real scalar field $\phi$ in $d+1$ dimensions with bulk euclidean action
\beq
\label{actionphi}
S^{(bulk)}[\phi] =  \int d^{d+1}x\,\sqrt{\det{g_{ab}}}\;
\left(\frac{1}{2}\,g^{ab}\;\partial_a\phi\; \partial_b\phi+U(\phi)\right)
\eeq
in a non-dynamical $AdS_{d+1}$ background
\beq\label{metric}
g = \frac{1}{z^2}\left(L^2dz^2+ \delta_{\mu\nu}\;dx^\mu\;dx^\nu\right)
\eeq
where $(x^\mu)$ are the QFT coordinates with $x^d\equiv i\,x^0$
the euclidean time, and $L$ the AdS scale.
The boundary is located at $z=0$ (UV region) while the horizon is at $z=\infty$
(IR region).

The equation of motion derived from (\ref{actionphi}) results,
\beq
\label{eomphi}
z^2\;\ddot{\phi}(x, z) -\left(d-1\right)z\; \dot{\phi}(x,z) + L^2\,z^2\,\Box\phi(x,z) =L^2\;U'(\phi)
\eeq
where $\Box\equiv \delta_{\mu\nu}\partial_\mu\partial_\nu$, and throughout the paper
we will indicate with a dot the derivative w.r.t. the bulk coordinate $z$ and with
a prime a field derivative.

We will consider potentials with
\beq\label{condpot}
U(0)=0\quad,\quad U'(0)=0\qquad;\qquad U(\phi_m)<0\quad,\quad U'(\phi_m)=0\quad,\quad U''(\phi_m)>0
\eeq
i.e. $\phi_m$ will be the true minimum, while at the origin the potential can
have a minimum (being a false vacuum thus) or even a maximum, provided that it is in the
Breitenl\"ohner-Freedman conformal window $-d^2/4<L^2\,U''(0)<0$.

It will be useful along the paper to work with dimensionless field variable
and potential defined by,
\beq\label{fieldpotredef}
t(z,x)\equiv\frac{\phi(z,x)}{\phi_m}\qquad;\qquad
V(t)\equiv \frac{L^2}{\phi_m^2}\;U(\phi_m\, t)
\eeq

We will be interested in regular, Poincar\`e invariant solutions $t=t(z)$
that interpolate between the UV and IR regions.
They obey the equation of motion
\beq\label{eom}
z^2\;\ddot{t}(z) -(d-1)\;z\; \dot{t}(z) = V'(t)
\eeq
and necessary behave in the UV and IR as
\beq\label{bc}
t(z)\xrightarrow{z\rightarrow 0} a_{UV}\; z^{\Delta^{UV}}
\qquad;\qquad t(z)\xrightarrow{z\rightarrow\infty} 1 + a_{IR}\; z^{d-\Delta^{IR}}
\eeq
respectively, where for the vev deformation case we will be considering
\beq\label{Delta}
\Delta^{UV/IR} \equiv \frac{d}{2} + \nu_{UV/IR}\qquad;\qquad
\nu_{UV/IR}\equiv\sqrt{\frac{d^2}{4} + m^2_{UV/IR}}
\eeq
with $m^2_{UV}\equiv V''(0)$ and $m^2_{IR}\equiv V''(1)>0$ ($t=1$ is a minimum according to
(\ref{condpot}))\footnote{\label{foot1}In the window $-\frac{d^2}{4}<V''(0)<0$ the term
$z^{d-\Delta^{UV}}$ could also be present in the small $z$ power expansion of $t(z)$.
From the AdS/CFT point of view this term is interpreted as a source that breaks
explicitly the scale invariance of the boundary QFT; then we should not expect a
Goldstone mode to appear, situation we are not interested in.
These domain walls are interpreted as dual to renormalization group flows generated by
deformation of the UV CFT by a relevant operator, i.e. one of dimension less than $d$ \cite{Skenderis:2002wp}.
}.

We recall as a last remark that the symmetries of $AdS$ space translate in the scale invariance of equation
(\ref{eom}), i.e. if $t(z)$ is a solution so it is $t(\lambda z)$, a fact of great relevance in what follows.

\subsection{\label{BPS}The BPS solutions}

The fact that the on-shell action vanishes\footnote{This could have been expected from the
spontaneous breaking of conformal symmetry, for a discussion on this point and references
see for example \cite{Berman:2002kd}.} \cite{Bajc:2012vk} is a hint that the
solution may be of the BPS type, i.e. it solves a first order equation.
Let us prove this statement in our context.
The action (\ref{actionphi}) is $\;S^{(bulk)}[\phi]= \frac{V_d\,\phi_m{}^2}{L}\,I[t]\;$,
with
\beq
I[t]=\int_0^\infty dz\; z^{-1-d}\,\left(\frac{1}{2}\,z^2\;\dot{t}(z)^2+V(t)\right)
\eeq
If we define the ``superpotential" $W(t)$ by,
\beq
\label{VwithW}
V(t)=\frac{1}{2}\,W'^2(t)-d\;W(t)
\eeq
then
\bea
\label{action}
\left.I[t]\right|_{on-shell}&=&\int_0^\infty dz\left[z^{-1-d}\frac{1}{2}\;\left(z\;\dot{t}(z) -W'(t)\right)^2+z^{-d}\;\dot{t}(z)\;W'(t)-d\;z^{-1-d}\;W(t)\right]\cr
&=&\int_0^\infty dz\;
\left[z^{-1-d}\;\frac{1}{2}\;\left(z\;\dot{t}(z) -W'(t)\right)^2
+z^{-d}\;\frac{dW(t)}{dz}+\frac{d}{dz}\left(z^{-d}\right)\;W(t)\right]\cr
&=&\int_0^\infty dz\;z^{-1-d}\;\frac{1}{2}\;\left(z\;\dot{t}(z)-W'(t)\right)^2
\eea
In the last line we used,
\beq
\left.\frac{W(t)}{z^d}\right|_{z=0}^{z=\infty}=
-\left.\frac{W(t)}{z^d}\right|_{z=0} = 0
\eeq
that follows from the fact that according to (\ref{bc}) $t(z)\sim z^{\Delta^{UV}}$ for ${z\to 0}$, and since $V(0)=V'(0)=0$
and $V''(0)$ finite, $W(t)\sim t^{2+n}\sim z^{n\Delta^{UV} + 2\nu_{UV}+d}$, with $n\ge0$.

It is now obvious from (\ref{action}) that there is a BPS like equation,
\beq
\label{eomBPS}
z\;\dot{t}(z)= W'(t(z))
\eeq
whose solutions satisfy also the full second order equation of motion (\ref{eom})
and for which the action (\ref{action}) vanishes.

Before analyzing some explicit examples we would like to notice the following relevant fact.
In the presence of dynamical gravity one must consider the gravity action
\beq\label{actiongrav}
S_{grav}[g] = -\frac{1}{16\,\pi\,G_N}\,\int d^{d+1}x\,\left(R[g] + \frac{d\,(d-1)}{L^2}\right)
\eeq
where $G_N$ is the $d+1$-dimensional Newton constant.
A Poincar\`e consistent ansatz that replaces (\ref{metric}) is,
\beq\label{dynmetric}
g = \frac{1}{z^2}\left(L^2\,\frac{dz^2}{F(z)}+ \delta_{\mu\nu}\;dx^\mu\;dx^\nu\right)
\eeq
If we now introduce the superpotential by replacing (\ref{VwithW}) with,
\beq\label{VwithWbr}
V(t)\equiv \frac{1}{2}\,W'^2(t)- d\,W(t) - \frac{d\,\kappa^2}{2}\;W^2(t)
\eeq
where $\kappa^2\equiv \frac{8\,\pi}{d-1}\phi_m{}^2\,G_N $, then (\ref{eomBPS}) 
gets
replaced by the following equations of motion,
\bea\label{eomBPSbr}
F(z)&=& \left(1 + \kappa^2\;W(t)\right)^2\\
z\;\dot{t}(z)&=& \frac{W'(t(z))}{1 + \kappa^2\,W(t)}
\eea
So, we conclude that there exist a smooth non-dynamical gravity $\kappa\to 0$ limit
which yields the system under consideration.

\subsubsection{\label{simplest}The simplest nontrivial example}

The simplest consistent nontrivial solution is for a quartic potential, i.e. a cubic ``superpotential" $W$.
The choice
\beq\label{W}
W(t)=\Delta\;\left(\frac{1}{2}\;t^2 - \frac{1}{3}\;t^3\right)\qquad;\qquad\Delta>\frac{d}{2}
\eeq
has the right properties (\ref{condpot}), i.e. $\,V(0)=V'(0)=V'(1)=0\,,\, V(1)=-\frac{d\,\Delta}{6}<0\,,\, V''(1)=\Delta\,(\Delta +d)>0$.
Furthermore $V''(0)=\Delta\,(\Delta -d)$ which implies that $t=0$ is a minimum
when $\Delta>d$ and a maximum if $0<\Delta<d$.
The solution to (\ref{eomBPS}) is,
\beq
\label{sol}
t(z)=\frac{z^\Delta}{1+z^\Delta}
\eeq
where we have fixed the scale invariance freedom.
On the other hand, the parameter $\Delta$  must be identified with $\Delta^{UV}$
in (\ref{Delta}) (or with $d-\Delta^{UV}$ if $0<\Delta<\frac{d}{2}$, but
we will not consider this case, see footnote on the previous page) while that in the IR region $z\to\infty$
the solution goes like $1-z^{d-\Delta^{IR}}$ with
\beq
\Delta^{IR}=d+\Delta
\eeq

\section{\label{methods}The propagator of the boundary theory}

Equation (\ref{eom}) is invariant under $z\to\lambda z$. A nontrivial domain wall solution at a vanishing source 
(vev deformation) breaks this invariance spontaneously, so one expects the appearance of a massless mode,
the Goldstone boson in the boundary theory \cite{Bianchi:2001de}. This was indeed confirmed in \cite{Bajc:2012vk}
for a piece-wise quadratic potential, solving exactly the equation of perturbations.

On the contrary, applying a matching method to find the solution at leading order in a small $q$ expansion 
gives a $1/q^{2\nu_{IR}}$ (but no $1/q^2$) term in the boundary field theory scalar propagator 
\cite{Hoyos:2012xc}. Puzzled by the discrepancy between \cite{Bianchi:2001de} and \cite{Hoyos:2012xc}
on the existence of the pole, we reanalyzed the problem here and we confirm the
usual $1/q^2$ pole behavior for the Goldstone as obtained in \cite{Bianchi:2001de}.

\subsection{\label{matching}The matching method}

We have seen many indications for the existence of the leading $1/q^2$ pole. So it seems that
there is something wrong with the matching method used in \cite{Hoyos:2012xc}, or at least
with the way it was implemented. What exactly went wrong will be shown in this section.

To compute the two-point function from holography we need to solve equation 
\beq\label{xieqbis}
z^2\;\ddot{\xi}(z;q)-(d-1)\;z\;\dot{\xi}(z;q)-\left(q^2\,z^2 + V''(t(z))\right)\;\xi(z;q)=0
\eeq
The idea is to match the large $z$ known solution of the form
\beq
\label{besselK}
\xi(z;q)\xrightarrow{z\to\infty} \xi_\infty(z;q) \equiv \frac{2}{\Gamma(\nu_{IR})}\;
\left(\frac{q}{2}\right)^{\nu_{IR}}\;z^\frac{d}{2}\;K_{\nu_{IR}}(q z)
\eeq
with some analytical solution of the perturbation equation for $q=0$.
Fortunately in the problem considered such solution of (\ref{xieqbis}) for $q=0$ is known.
As noted in Subsection 2.4 due to dilatation invariance of
the equation of motion (e.o.m.), one solution is
\beq\label{xi+}
\xi_+(z)=z\; \dot{t}(z)
\eeq
with $t(z)$ the solution of the e.o.m., while the second can be found from the integral
\beq\label{xi-}
\xi_-(z)=\xi_+(z)\left(\int_{z_i}^z dy\frac{y^{d-1}}{\xi_+^2(y)}+\frac{\xi_-(z_i)}{\xi_+(z_i)}\right)
\eeq
where $z_i$ and $\xi_-(z_i)$ are integration constants; of course the physics can not depend on the choice of them.

The matching method at the leading order in $q$ 
consists of merging at some large $z$, but small $qz$, the two solutions,
%The idea is to match at some large $z$, but small $qz$, the two solutions,
i.e. determine the ratio $C_+(q)/C_-(q)$ from the behavior of the leading
$z^{d-\Delta^{IR}}$ and $z^{\Delta^{IR}}$ terms of (\ref{besselK}) and
the approximate solution
\beq
\label{anyz}
\xi(z;q)\approx C_+(q)\;\xi_+(z)+C_-(q)\;\xi_-(z)
\eeq
Since from their definitions (\ref{xi+}) and (\ref{xi-}) the two solutions go in the UV  as
\beq
\xi_+(z)\xrightarrow{z\to 0} \Delta^{UV}\;z^{\Delta^{UV}}\qquad;\qquad
\xi_-(z)\xrightarrow{z\to 0} \frac{z^{d-\Delta^{UV}}}{(d-2\,\Delta^{UV})\,\Delta^{UV}}
\eeq
the boundary propagator can be approximately calculated for $q\to 0$ from
\beq
\label{propagator}
G_2(q)=\left(\Delta^{UV}\right)^2\; \left(d- 2\,\Delta^{UV}\right)\;\frac{C_+(q)}{C_-(q)}
\eeq
The existence of an overlapping region has been proved in \cite{Hoyos:2012xc}, where it
was also shown that the dominant behavior of the propagator (\ref{propagator}) is 
$1/q^{2\nu_{IR}}$, see (\ref{propagatorwrong}).

The method implicitly assumes that the approximation (\ref{anyz}) is good enough.
We will show now that this is not the case for the issue of the propagator. Let's see
what happens at the next order in $q^2$. The solution gets expanded as power series in
$q^2$:
\beq
\xi(z;q)=\sum_{n=0}^\infty q^{2n}\;\xi^{(n)}(z;q)
\eeq
where now (\ref{anyz}) is just the leading order
\beq
\xi^{(0)}(z;q)=C_+^{(0)}(q)\;\xi_+(z)+C_-^{(0)}(q)\;\xi_-(z)
\eeq
and the $n-$th term solves
\beq
\label{xin}
z^2\;\ddot{\xi}^{(n)}(z;q)-(d-1)\;z\;\dot{\xi}^{(n)}(z;q)-
V''(t(z))\;\xi^{(n)}(z;q)=z^2\;\xi^{(n-1)}(z;q)\hskip1cm n=1,2,\ldots
\eeq
Let us solve it for $n=1$:
\bea
\xi^{(1)}(z;q)&=&C_+^{(1)}(q)\;\xi_+(z)+C_-^{(1)}(q)\;\xi_-(z)\cr
&+&C_+^{(0)}(q)\;\xi_+(z)\,\int_{z_i}^z dx\,\frac{x^{d-1}}{\xi_+^2(x)}\int_{z_i}^x dy\,\frac{\xi_+^2(y)}{y^{d-1}}\cr
&+&C_-^{(0)}(q)\;\xi_-(z)\,\int_{z_i}^z dx\,\frac{x^{d-1}}{\xi_-^2(x)}\int_{z_i}^x dy\,\frac{\xi_-^2(y)}{y^{d-1}}
\eea
where the first two terms on the right-hand-side represent the general solution of the
homogeneous equation, while the last two terms are particular solutions of the
non-homogeneous equation. In deriving it
we took into account the linearity of (\ref{xin}).

The next-to-leading order solution is thus
\bea
\xi(z;q)&\approx& \xi_+(z)\;\left[\left(C_+^{(0)}(q)+q^2\,C_+^{(1)}(q)\right)+
q^2\,C_+^{(0)}(q)\,\int_{z_i}^z dx\,\frac{x^{d-1}}{\xi_+^2(x)}\int_{z_i}^x dy\,\frac{\xi_+^2(y)}{y^{d-1}}\right]\cr
&+& \xi_-(z)\;\left[\left(C_-^{(0)}(q)+q^2\,C_-^{(1)}(q)\right)+
q^2\,C_-^{(0)}(q)\,\int_{z_i}^z dx\,\frac{x^{d-1}}{\xi_-^2(x)}\int_{z_i}^x dy\,\frac{\xi_-^2(y)}{y^{d-1}}\right]\cr
&&
\eea
At this order of the $q^2$ expansion we can replace
\beq
q^2\,C_\pm^{(0)}\to q^2\,\left(C_\pm^{(0)}(q)+q^2\,C_\pm^{(1)}(q)\right)
\eeq
So we are left with just two integration constants, denoted from now on by
\beq
C_\pm(q)\equiv C_\pm^{(0)}(q)+q^2\,C_\pm^{(1)}(q)
\eeq
as it should be for a second order differential equation.
The solution is now
\bea
\label{xi}
\xi(z;q)&\approx& C_+(q)\;\xi_+(z)\left(1+q^2
\int_{z_i}^z dx\,\frac{x^{d-1}}{\xi_+^2(x)}\int_{z_i}^x dy\,\frac{\xi_+^2(y)}{y^{d-1}}\right)\nonumber\\
&+& C_-(q)\;\xi_-(z)\left(1+q^2
\int_{z_i}^z dx\,\frac{x^{d-1}}{\xi_-^2(x)}\int_{z_i}^x dy\,\frac{\xi_-^2(y)}{y^{d-1}}\right)
\eea

Finally, for later use, the double integral can be simplified as
\beq
\xi_\pm(z)\int_{z_i}^z dx\frac{x^{d-1}}{\xi_\pm^2(x)}\int_{z_i}^x dy\frac{\xi_\pm^2(y)}{y^{d-1}}=
\pm\xi_\mp(z)\int_{z_i}^z dx\frac{\xi_\pm^2(x)}{x^{d-1}}\mp
\xi_\pm(z)\int_{z_i}^z dx\frac{\xi_+(x)\xi_-(x)}{x^{d-1}}
\eeq
where we used the relation
\beq
\frac{z^{d-1}}{\xi_\pm^2(z)}=\pm\left(\frac{\xi_\mp(z)}{\xi_\pm(z)}\right)'
\eeq
that follows from the definitions (\ref{xi+}), (\ref{xi-}).

\subsubsection{\label{explicit}An explicit calculation}

A full analysis for an arbitrary superpotential is rather cumbersome. 
Instead of presenting it we prefer to focus on the particular case of 
subsection \ref{simplest}. The extension to a generic superpotential 
is straightforward.

The solution to the equation of motion is 
\beq
\label{tz}
t(z)=\frac{z^\Delta}{1+z^\Delta}
\eeq
The corresponding solutions to the equation of perturbation at $q=0$ are
\bea
\label{xi+exp}
\xi_+(z)&=&\frac{\Delta z^\Delta}{(1+z^\Delta)^2}\\
\label{xi-exp}
\xi_-(z)&=&\frac{1}{\Delta(1+z^\Delta)^2}\sum_{k=0}^{4}
\bem
4 \\ k
\eem
\frac{z^{d+(k-1)\Delta}}{d+(k-2)\Delta}
\eea
with the limits,
\bea\label{zto0}
\xi_+(z)\xrightarrow{z\to 0} a_+^{UV}\;z^\Delta\qquad&;&\qquad
\xi_+(z)\xrightarrow{z\to\infty} a_+^{IR}\;z^{-\Delta}\\
\label{ztoinfty}
\xi_-(z)\xrightarrow{z\to 0} a_-^{UV}\;z^{d-\Delta}\qquad&;&\qquad
\xi_-(z)\xrightarrow{z\to\infty} a_-^{IR}\;z^{d+\Delta}
\eea
where the explicit values of the constants are given in Table \ref{tab1}.
\begin{table}
\centering
 \begin{tabular}{| c || c | c |}
   \hline
        & $a_+$    & $a_-$    \\ \hline\hline
  UV & $\Delta$  & $\frac{1}{\Delta(d-2\Delta)}$  \\ \hline
  IR   & $\Delta$ & $\frac{1}{\Delta(d+2\Delta)}$   \\ \hline
 \end{tabular}
 \caption{The explicit values in (\ref{zto0}) and (\ref{ztoinfty}).}
 \label{tab1}
\end{table}
We find
\bea
\int_{z_i}^z dx\frac{\xi_+^2(x)}{x^{d-1}}&=&\Delta \;B_{(t,t_i)}\left(2-\frac{d-2}{\Delta},2+\frac{d-2}{\Delta}\right)\\
\int_{z_i}^z dx\frac{\xi_+(z)\xi_-(x)}{x^{d-1}}&=&\frac{1}{\Delta}\sum_{k=0}^{4}\frac{1}{d+(k-2)\Delta}
\begin{pmatrix}
4 \\ k
\end{pmatrix}
B_{(t,t_i)}\left(\frac{2}{\Delta}+k,4-\frac{2}{\Delta}-k\right)\\
\int_{z_i}^z dx\frac{\xi_-^2(x)}{x^{d-1}}&=&\frac{1}{\Delta^3}\sum_{k,l=0}^{4}\frac{1}{d+(k-2)\Delta}
\begin{pmatrix}
4 \\ k
\end{pmatrix}
\frac{1}{d+(l-2)\Delta}
\begin{pmatrix}
4 \\ l
\end{pmatrix}\nonumber\\
&\times&B_{(t,t_i)}\left(\frac{d+2}{\Delta}+k+l-2,-\frac{d+2}{\Delta}-k-l+6\right)
\eea
where $t=t(z)$, $t_i=t(z_i)$ and
\beq
B_{(t,t_i)}(a,b)\equiv\int_{t_i}^t d\tau\tau^{a-1}(1-\tau)^{b-1}
\eeq
is the generalized incomplete (Euler) beta function.
They can be expanded as
\bea
{\rm IR}\;(t\to 1)&:&B_{(t,t_i)}(a,b)=-\sum_{n=0}^\infty\frac{(-1)^n\Gamma(a)
\left((1-t)^{b+n}-(1-t_i)^{b+n}\right)}{(b+n)\Gamma(n+1)\Gamma(a-n)}\\
{\rm UV}\;(t\to 0)&:&B_{(t,t_i)}(a,b)=\sum_{n=0}^\infty\frac{(-1)^n\Gamma(b)
\left(t^{a+n}-t_i^{a+n}\right)}{(a+n)\Gamma(n+1)\Gamma(b-n)}
\eea
Now we can combine the different expansions to get the needed powers:
\bea
\xi_+(z)&\to&a_+^{IR} z^{-\Delta}\\
\xi_-(z)&\to&a_-^{IR} z^{d+\Delta}\\
\xi_+(z)\int_{z_i}^z dx\frac{x^{d-1}}{\xi_+^2(x)}\int_{z_i}^x dy\frac{\xi_+^2(y)}{y^{d-1}}&\to&
-\epsilon_{+-}^{IR}a_+^{IR} z^{-\Delta}+\epsilon_{++}^{IR}a_-^{IR} z^{d+\Delta}\\
\xi_-(z)\int_{z_i}^z dx\frac{x^{d-1}}{\xi_-^2(x)}\int_{z_i}^x dy\frac{\xi_-^2(y)}{y^{d-1}}&\to&
-\epsilon_{--}^{IR} a_+^{IR}z^{-\Delta}+\epsilon_{+-}^{IR}a_-^{IR} z^{d+\Delta}
\eea
with
\bea
\epsilon_{++}^{IR}&=&
\Delta B_{(1,t_i)}(2-(d-2)/\Delta,2+(d-2)/\Delta)\\
\epsilon_{+-}^{IR}&=&
\sum_{k=0}^{4}
\begin{pmatrix}
4 \\ k
\end{pmatrix}
\frac{B_{(1,t_i)}(2/\Delta+k,4-2/\Delta-k)}{\Delta(d+(k-2)\Delta)}\\
\epsilon_{--}^{IR}&=&\sum_{k,l=0}^{4}\frac{1}{d+(k-2)\Delta}
\begin{pmatrix}
4 \\ k
\end{pmatrix}
\frac{1}{d+(l-2)\Delta}
\begin{pmatrix}
4 \\ l
\end{pmatrix}\cr
&\times&\frac{B_{(1,t_i)}(-2+(d+2)/\Delta+k+l,6-(d+2)/\Delta-k-l)}{\Delta^3}
\eea
where we used
\beq
B_{(t_2,t_1)}(a,b)=B_{(1-t_1,1-t_2)}(b,a)
\eeq

We get close to $z=\infty$
\bea
\xi(z;q)&\to&
\left[\left(1-q^2\epsilon_{+-}^{IR}\right)C_+(q)-
q^2\epsilon_{--}^{IR} C_-(q)\right]a_+^{IR}z^{-\Delta}\nonumber\\
&+&\left[\left(1+q^2\epsilon_{+-}^{IR}\right)C_-(q)+
q^2\epsilon_{++}^{IR} C_+(q)\right]a_-^{IR}z^{d+\Delta}
\eea

The system to solve after matching to (\ref{besselK}) is
\bea
\label{matching1}
a_+^{IR}\left[\left(1-q^2\epsilon_{+-}^{IR}\right)C_+(q)+
q^2\epsilon_{--}^{IR} C_-(q)\right]&=&1\\
\label{matching2}
a_-^{IR}\left[\left(1+q^2\epsilon_{+-}^{IR}\right)C_-(q)+
q^2\epsilon_{++}^{IR} C_+(q)\right]&=&\gamma q^{2\nu_{IR}}
\eea
where
\beq
\label{gamma}
\gamma\equiv \frac{\Gamma(-\nu_{IR})\,}{2^{2\,\nu_{IR}}\,\Gamma(\nu_{IR})}
\eeq
and where in this case
\beq
\nu_{IR}=\Delta+d/2
\eeq

The leading order solution for small $q$ is
\bea
\label{C+}
C_+(q)&=&\frac{1}{a_+^{IR}}+\ldots\\
\label{C-}
C_-(q)&=&-\epsilon_{++}^{IR}q^2C_+(q)+\ldots=
-\frac{\epsilon_{++}^{IR}}{a_+^{IR}}q^2+\ldots
\eea
If we keep only the formally leading order expression, i.e. no $\epsilon$'s in
(\ref{matching1})-(\ref{matching2}), we would get a wrong result: although $C_+(q)$ remains the same,
$C_-(q)=(\gamma/a_-^{IR})q^{2\nu_{IR}}$ changes drastically.

For $z\to 0$ on the other side
\bea
\xi_+(z)&\to&a_+^{UV}z^\Delta\\
\xi_-(z)&\to&a_-^{UV}z^{d-\Delta}\\
\xi_+(z)\int_{z_i}^z dx\frac{x^{d-1}}{\xi_+^2(x)}\int_{z_i}^x dy\frac{\xi_+^2(y)}{y^{d-1}}&\to&
-\epsilon_{+-}^{UV}a_+^{UV}z^\Delta+\epsilon_{++}^{UV}a_-^{UV}z^{d-\Delta}\\
\xi_-(z)\int_{z_i}^z dx\frac{x^{d-1}}{\xi_-^2(x)}\int_{z_i}^x dy\frac{\xi_-^2(y)}{y^{d-1}}&\to&
-\epsilon_{--}^{UV}a_+^{UV}z^\Delta+\epsilon_{+-}^{UV}a_-^{UV}z^{d-\Delta}
\eea
with
\bea
\epsilon_{++}^{UV}&=&-\Delta B_{(t_i,0)}(2-(d-2)/\Delta,2+(d-2)/\Delta)\\
\epsilon_{+-}^{UV}&=&
-\sum_{k=0}^{4}
\begin{pmatrix}
4 \\ k
\end{pmatrix}
\frac{B_{(t_i,0)}(2/\Delta+k,4-2/\Delta-k)}{\Delta(d+(k-2)\Delta)}\\
\epsilon_{--}^{UV}&=&-\sum_{k,l=0}^{4}\frac{1}{d+(k-2)\Delta}
\begin{pmatrix}
4 \\ k
\end{pmatrix}
\frac{1}{d+(l-2)\Delta}
\begin{pmatrix}
4 \\ l
\end{pmatrix}\cr
&\times&\frac{B_{(t_i,0)}(-2+(d+2)/\Delta+k+l,6-(d+2)/\Delta-k-l)}{\Delta^3}
\eea
so that
\bea
\xi(z;q)&\to&\left[\left(1-q^2\epsilon_{+-}^{UV}\right)C_+(q)-
q^2\epsilon_{--}^{UV}C_-(q)\right]a_+^{UV}z^\Delta\\
&+&\left[\left(1+q^2\epsilon_{+-}^{UV}\right)C_-(q)
+q^2\epsilon_{++}^{UV}C_+(q)\right]a_-^{UV}z^{d-\Delta}\nonumber
\eea
The propagator is then,
\beq
\label{g2qpartial}
G_2(q)\approx
\frac{a_+^{UV}\left[\left(1-q^2\epsilon_{+-}^{UV}\right)C_+(q)-
q^2\epsilon_{--}^{UV}C_-(q)\right]}
{a_-^{UV}\left[\left(1+q^2\epsilon_{+-}^{UV}\right)C_-(q)
+q^2\epsilon_{++}^{UV}C_+(q)\right]}
\xrightarrow{q^2\to0}\frac{\alpha}{q^2}
\eeq
where in general one can prove that
\beq
\alpha=\frac{a_+^{UV}/a_-^{UV}}{\epsilon_{++}^{UV}-\epsilon_{++}^{IR}}=
\frac{2\,\nu_{UV}\,(a_+^{UV})^2}{\int_{0}^\infty dw\;w^{1-d}\; \xi_+(w)^2}>0
\eeq
Eq. (\ref{g2qpartial}) confirms the Goldstone theorem. 
In our concrete example this gives
\beq
\alpha=\frac{\Delta(2\Delta-d)}{B_{(1,0)}(2-(d-2)/\Delta,2+(d-2)/\Delta)}
\eeq
The final result is independent on the arbitrary parameter $t_i=t(z_i)$, as it should be,
although several intermediate quantities depend on it. 

On the other side, if we retain only the formally leading order in $q^2$, 
we would get a wrong limit for the propagator,
\beq
\label{propagatorwrong}
G_2(q)\approx \left[(a_+^{UV}/a_-^{UV})(a_-^{IR}/a_+^{IR})
(1/\gamma)\right]q^{-2\nu_{IR}}
\eeq

One last comment: using the leading order expansion in small $q$ and the values (\ref{C+})-(\ref{C-}) 
it is easy to check explicitly that the zero mode solution is proportional to $\xi_+(z)$ (\ref{xi+exp}) and is thus 
normalizable\footnote{We thank the referee for this remark.} \cite{Berg:2006xy}.

\subsubsection{Beyond $q\to 0$}

What we derived is strictly speaking valid exactly only for $q\to 0$.
For all $z$ we can then use the solution for
infinitely small $q$, i.e. (\ref{xi}).
What if $q$ is small but non-zero? Can we use the same solution (\ref{xi})?

The point is that for a finite $q$ even matching must be done at a finite $z$.
In fact one must find a region in the $z-q$ plane where both approximations are valid.
More precisely, for large enough $z>z_\infty$ (see below) the small $q$ approximate solution
(\ref{xi}) (or a better one) is valid for $z\lsim z_0(q)$ where\footnote{except
for small regions in which $V''(t(z))\approx 0$.}
\beq
z_0(q)\equiv|V''(1)|^{1/2}/q
\eeq
while the large $z$ solution (\ref{besselK}) (or a better one) is valid
for $z\gsim z_\infty$:

\beq
z_\infty\equiv\left|\frac{V''(1)}{V'''(1)\;a_{IR}}\right|^\frac{1}{\Delta^{IR}}
\eeq
Then one has to match the two solutions not at $z\to\infty$ as we did in the previous example but at
a finite although large enough $z_\infty\lsim z\lsim z_0(q)$, which may modify the values of $C_\pm(q)$.
In order for such a region to exist at all, $z_\infty<z_0(q)$, which gives an upper bound for
$q$ from the solution of $z_\infty=z_0(q_{max})$. For higher $q$ the method fails.

\begin{figure}[htb]
\begin{center}
\includegraphics[width=10.cm]{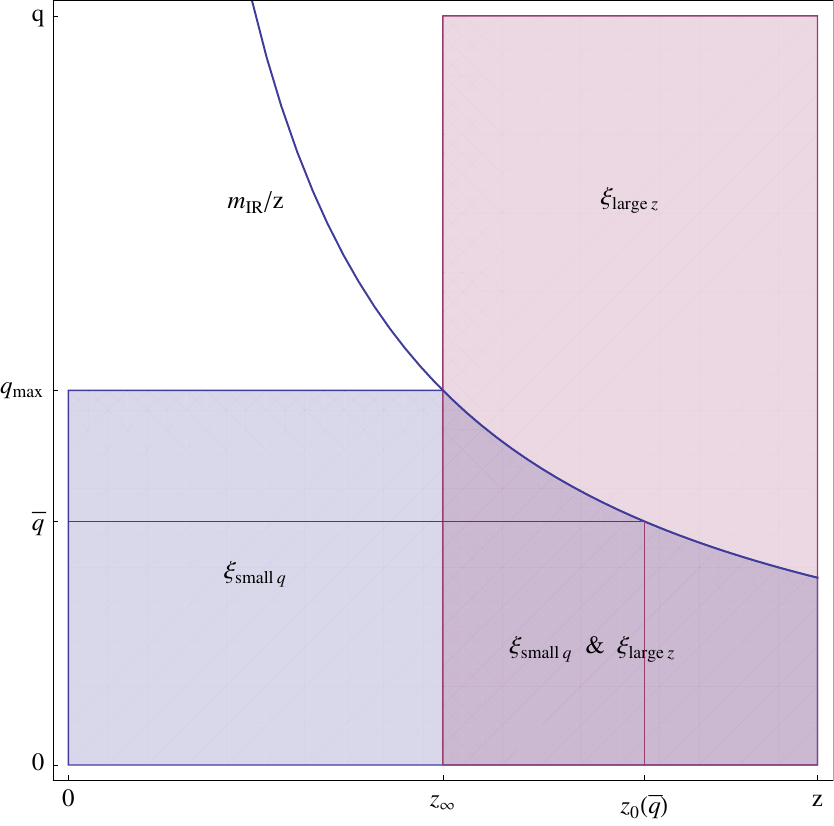}
\caption{\label{zq} A schematic diagram of the $z$-$q$ plane divided into regions
in which different approximations are valid. $\xi_{large\;z}$ is approximated by (\ref{besselK}),
$\xi_{small\;q}$ by (\ref{xi}), and $m_{IR}\equiv|V''(1)|^{1/2}$. For any $\bar q\leq q_{max}$ the same
approximate solutions can be used. In the white region all approximations mentioned in this paper fail.
}
\end{center}
\end{figure}

So the matching depends on $q$, although once we match for a given $q$, let us call it $\bar q$, then it
is valid for all $q\leq\bar q$. Then we can use the solution (\ref{xi}) (or a better one) for all $z\lsim z_0(\bar q)$ and the
solution (\ref{besselK}) (or a better one) for all $z\gsim z_\infty$. The situation is summarized on fig. \ref{zq}.

\section{Conclusions and outlook}

We considered in this work the bulk system of a real scalar field in a non-dynamical
AdS background. After finding BPS-type solutions of the bulk equation of motion,
we solved the perturbation equation on this background in two different limits,
the large $z$ and the small $q$ regimes. We have shown that a correct application of 
the matching procedure between these two approximate solutions leads to a massless 
pole, as expected by the Goldstone theorem applied to the spontaneous breaking of 
dilatation invariance \cite{Bianchi:2001de}.

Is the result of the simple pole propagator in the boundary theory a consequence of the
no-backreaction ($\kappa\to 0$) limit?
We believe that this is not the case, and that the result is generic.
In fact, neither at finite nor at vanishing $\kappa$ the formally leading order
of the matching method gives the $1/q^2$ behavior, which indicates that the
problem is in the application of the method and not in the difference of the systems.

As we saw, given a potential, there is a maximal momentum for which we can apply 
the matching method. The result is in the form of a positive power expansion in $q$. The 
limitation is due to the vanishing of a common region for the low $q$ and large $z$
expansions.
It is thus possible that an analytic continuation of the result exists for all $q$,
although, due to the perturbative character of the solution, there is little hope to find it. 
%A study of finite $q^2$ propagator would be interesting to find out possible other
%poles (and thus bound states in the boundary theory) of the propagator \cite{Berg:2006xy}.

Within the approximations of this paper we can calculate the on-shell 1-particle irreducible 3-point
correlator. We find it zero, as required for a Goldstone boson.

Finally, there have been several discussions on the role of the a-theorem
\cite{Komargodski:2011vj,Komargodski:2011xv,Luty:2012ww} in holography
(see for example \cite{Myers:2010tj}, \cite{Hoyos:2012xc} and references therein).
The value (especially the positivity) of the ${\cal O}(q^4)$ coefficient \cite{Komargodski:2011vj}
in the $2\to2$ dilaton scattering amplitude represents a check of the AdS/CFT correspondence
in connection with the a-theorem. Unfortunately, the calculation explicitly involves
the bulk-bulk propagator, which we miss at the moment. We leave this interesting issue for the future.

\section*{Acknowledgments}
We would like to thank Mirjam Cveti\v c, Fidel Schaposnik and Guillermo Silva for discussions,
and Mirjam Cveti\v c, Carlos Hoyos, Uri Kol, Cobi Sonnenschein and Shimon Yankielowicz
for correspondence. This work has been supported in part by the Slovenian Research
Agency, and by the Argentinian-Slovenian programme BI-AR/12-14-004 //
MINCYT-MHEST SLO/11/04.

\end{document}